\documentclass[10pt,a4paper]{article}
\usepackage{amssymb}
\usepackage{amsmath}
\usepackage{amsfonts}
\usepackage{amsthm}
\usepackage{latexsym}
\usepackage{graphicx}                   
\usepackage{epstopdf}                   
\usepackage[dvips]{epsfig}
\usepackage{setspace}
\usepackage{float}
\usepackage{wrapfig}
\usepackage{bm}
\usepackage{enumerate}
\usepackage{authblk}
\usepackage{mathtools}
\usepackage{color}                          
\usepackage{soul}                           
\usepackage{nicefrac}                     

\usepackage{caption}
\usepackage{subcaption}
\begin{document}

\title{\textbf{A Corkscrew Model for Highly Coupled Anisotropic Compliance in Ruddlesden--Popper Oxides with Frozen Octahedral Rotations}}

\author[,1]{Chris Ablitt}
\affil[1]{Department of Materials, Imperial College London, London SW7 2AZ, U.K.}
\author[,2]{Mark S. Senn \footnote{E-mail address: {\tt m.senn@warwick.ac.uk}}}
\affil[2]{Department of Chemistry, University of Warwick, Gibbet Hill, Coventry, CV4 7AL, U.K.}
\author[,3]{Nicholas C. Bristowe \footnote{E-mail address: {\tt n.c.bristowe@kent.ac.uk}}}
\affil[3]{School of Physical Sciences, University of Kent, Canterbury CT2 7NH, U.K.}
\author[,1,4]{Arash A. Mostofi}
\affil[4]{Department of Physics, Imperial College London, London SW7 2AZ, U.K.}
\date{\today}

\maketitle

\begin{abstract}
A ``corkscrew" mechanism, that couples changes in the in-plane rotation angle to strains along the layering axis, has been proposed previously to explain increased compliance in certain Ruddlesden--Popper phases that facilitates uniaxial negative thermal expansion over a wide temperature range. Following the procedure developed to study many simple, auxetic geometries, in the present study we derive the elastic compliances predicted by this corkscrew mechanism assuming that the four shortest metal--anion bonds remain stiff and changes in bond angle are modelled by a harmonic angle potential. We subsequently analyse the limitations of this model and show that it may be extended to A$_{n+1}$B$_n$O$_{3n+1}$ Ruddlesden--Popper oxide phases of general layer thickness $n$.

\end{abstract}

\section{Introduction}

The study of \emph{negative} material properties, such as \emph{negative Poisson's ratio} (NPR), \emph{negative linear compressibility} (NLC) and \emph{negative thermal expansion} (NTE), has become an exciting field in materials chemistry. These unusual properties defy conventional intuition regarding how materials should behave and as such much work has gone into developing mechanisms to explain their occurrence in the rare examples of materials in which these phenomena manifest. In many cases, especially in metal organic framework (MOF) and inorganic framework materials, these explanations have involved describing the structure using simple geometric models.

The $Acaa$ phase of Ca$_3$Mn$_2$O$_7$ exhibits uniaxial negative thermal expansion over a wide temperature range of approximately 950 K -- between when the NTE phase first coexists alongside a competing low temperature phase up until when the material decomposes \cite{Senn2015}. This compound is a member of the Ruddlesden-Popper (RP) oxide series, a class of layered perovskite materials with general formula A$_{n+1}$B$_n$O$_{3n+1}$ where A and B are cations and $n$ denotes the number of ABO$_3$ perovskite layers stacked along the $c$ axis, with blocks of $n$ ABO$_3$ layers separated by a single AO rock salt layer. This discovery of NTE accompanies observations of uniaxial NTE along the layering axis in the analogous $n=1$ $I4_1/acd$ phase of Ca$_2$MnO$_4$ \cite{Takahashi1993Ca2MnO4, Ablitt2017}, Sr$_2$RhO$_4$~\cite{Huang1994, VogtButtrey1996, Ranjbar2015} and Sr$_2$IrO$_4$~\cite{Huang1994} where, as in $n=2$ $Acaa$ Ca$_3$Mn$_2$O$_7$, rotations of BO$_6$ octahedra about the layering axis are frozen into the structure. RP phases with frozen octahedral rotations about the layering axis (but no frozen octahedral tilts about an in-plane axis) will herein be referred to as rotation phases.

NTE is often explained in similar framework materials using the theory of rigid unit modes \cite{Dove1995, DoveReview}, soft vibrations of approximately rigid polyhedral structural units that drive contraction with increased temperature. Although we linked the decrease in magnitude of uniaxial NTE as $x$ increases in Ca$_{3-x}$Sr$_x$Mn$_2$O$_7$ with the hardening of low frequency tilts (with tilt axis lying in the layer plane) of octahedra in DFT calculations \cite{Senn2016}, in a computational study performed on $I4_1/acd$ Ca$_2$GeO$_4$ we showed that highly anisotropic compliance is an essential ingredient for anisotropic NTE in these materials, alongside the soft phonons that provide the dynamic driving force \cite{Ablitt2017}. The thermodynamic formalism distinguishing the vibrational contributions to thermal expansion from elements of the elastic compliance tensor was first derived almost 100 years ago \cite{GruneisenGoens1924} and the idea that anisotropic thermal expansion could be caused by anisotropic compliance was discussed based on experimental measurements of simple elements 40 years later \cite{BarronMunn1967}. In more recent years, studies based on first principles calculations have also discussed phonons and compliance separately when explaining computed thermal expansion \cite{FangDoveRefsonAg3CoCN62013}.

\begin{figure}[h!]
\centering
\begin{subfigure}{\textwidth}
\centering
\caption{}
\label{fig:corkscrewmechanism}
\includegraphics[width=0.95\textwidth]{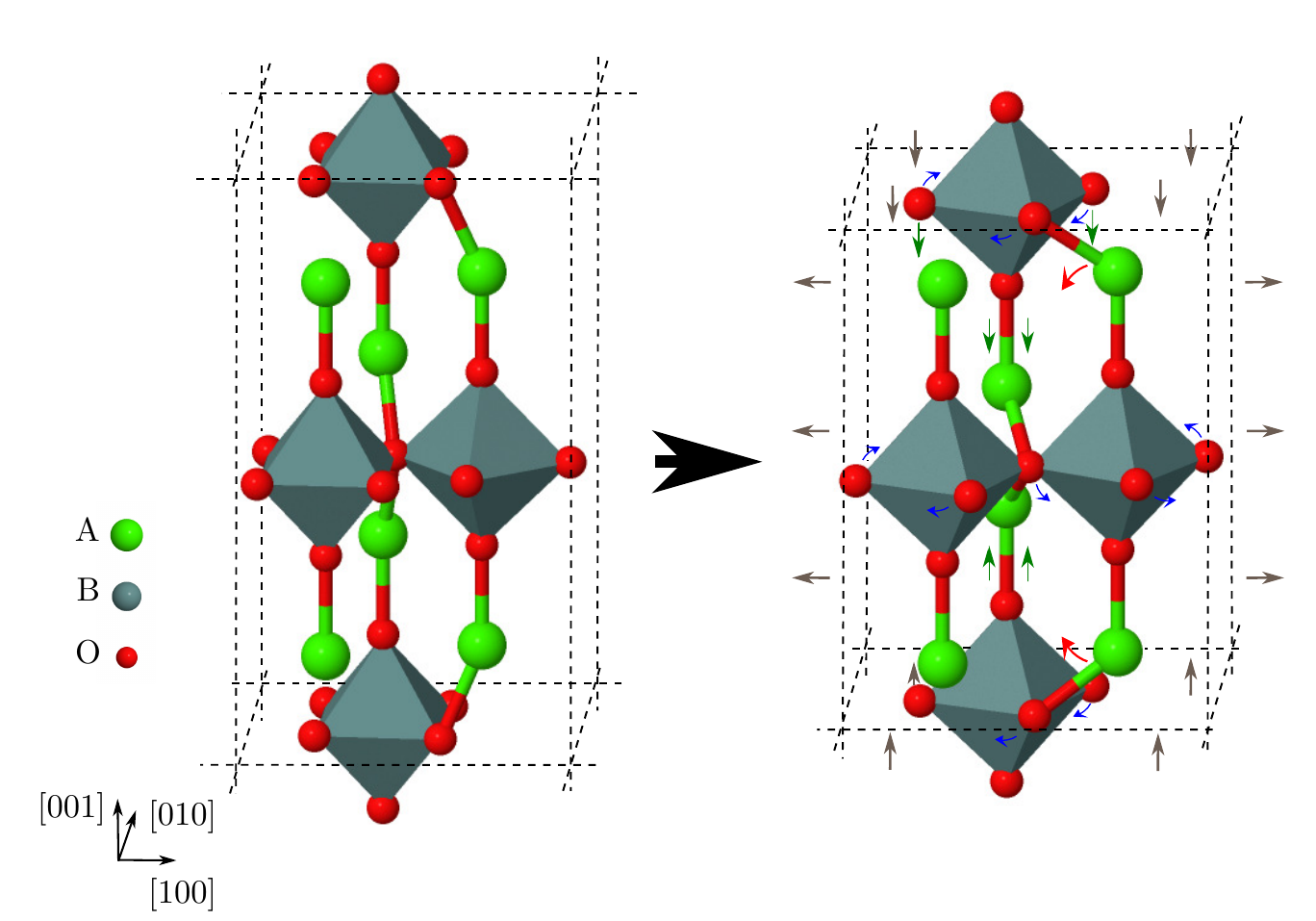}
\end{subfigure}\\[0.5cm]
\begin{subfigure}{0.45\textwidth}
\centering
\caption{}
\label{fig:realcorkscrew}
\includegraphics[width=0.95\textwidth]{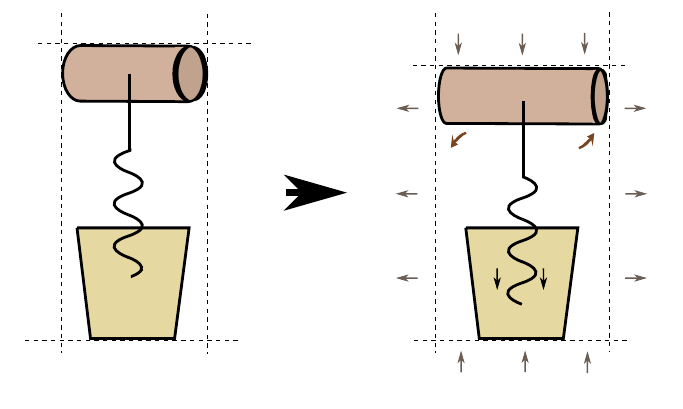}
\end{subfigure}
\hspace{0.5cm}
\begin{subfigure}{0.45\textwidth}
\centering
\caption{}
\label{fig:squares}
\includegraphics[width=0.95\textwidth]{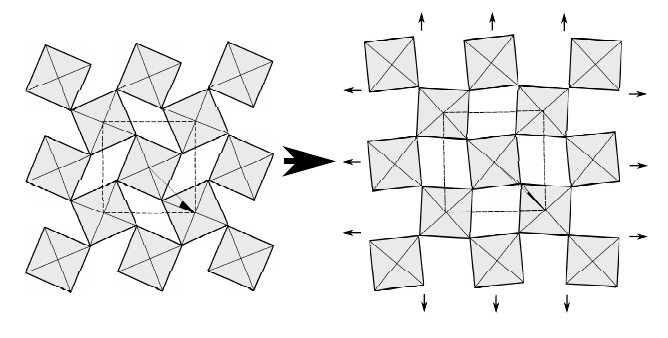}
\end{subfigure}\\[0.1cm]
\caption{{\footnotesize(a)} The \emph{corkscrew} mechanism in A$_2$BO$_4$ Ruddlesden--Popper oxides with a frozen rotation about the layering axis: a decrease in the in-plane rotation angle causes an in-plane expansion and a contraction along the layering axis; {\footnotesize(b)} a corkscrew being screwed into a cork: a change in the rotation angle also leads to an increase in the width of the two-dimensional picture but a decrease in the height of the combined object; and {\footnotesize(c)} the two-dimensional unit cell of rotated rigid corner-connected squares increases with a decrease in the rotation angle of each square.}
\label{fig:illustration}
\end{figure}

In order to answer the question ``why is NTE often seen in Ruddlesden-Popper oxide phases with a frozen octahedral rotation and not in equivalent phases of ABO$_3$ perovskite?" we proposed an atomic mechanism to explain this compliance that operates at the layer interface of RP rotation phases \cite{Ablitt2017}. Since neither RP and ABO$_3$ rotation phases have frozen octahedral tilts, it is likely that these modes will be active to provide a thermodynamic driving force for uniaxial NTE in both structures and therefore the difference in anisotropic compliance is a key factor distinguishing the likelihood of the two materials to exhibit uniaxial NTE. Assuming that in a rotation phase of A$_2$BO$_4$, the four most stiff metal -- oxygen bonds remain rigid (the two distinct intra-octahedral B--O bonds and the two shortest A--O bonds), it is possible to cooperatively expand the $a$ and $b$ in-plane lattice parameters and contract the $c$ layering axis by changing only bond angles allowed by the symmetry operations of the phase. This mechanism has been illustrated in Figure  \ref{fig:corkscrewmechanism}. We liken this mechanism to a \emph{corkscrew} since, like a corkscrew being screwed into a cork, an in-plane rotation leads to a decrease in height of the combined object (Figure \ref{fig:realcorkscrew}). In the layering plane, the three-dimensional BO$_6$ corner-linked octahedra are viewed as two-dimensional squares and therefore it should be clear from Figure \ref{fig:squares} that changing the rotation angle of these rigid squares leads to a change in the in-plane lattice parameter. 

In this paper we analyse this \emph{corkscrew} mechanism and derive the elastic compliance matrix predicted if these bonds do indeed remain stiff and the resistance to deformation comes from a harmonic potential on certain metal--oxygen--metal bond angles. This follows the method used to analyse similar geometric systems, often in the field of auxetic (NPR) materials, for example to study simple two-dimensional systems formed from corner connected squares \cite{Smith2000}, triangles \cite{Grimar2011Triangles}, rhombi \cite{Grima2008Rhombi} or rectangles of different sizes \cite{Grima2010SquaresandRectangles} or even to study more complex three-dimensional systems \cite{Grima2012} as the RP structure is. Although we do not expect idealised models of this kind to represent exactly real chemical systems, analysis of this kind can be useful to understand mechanisms that operate in a real material alongside other effects. This manuscript is intended to support our recently accepted publication \cite{AblittFrontiers2018} where we assess through first-principles calculations how the anisotropic compliance changes in RP compounds with changing layer thickness $n$.

\section{Derivation}
\label{sec:Derivation}

\begin{figure}[h!]
\includegraphics[width=0.95\textwidth]{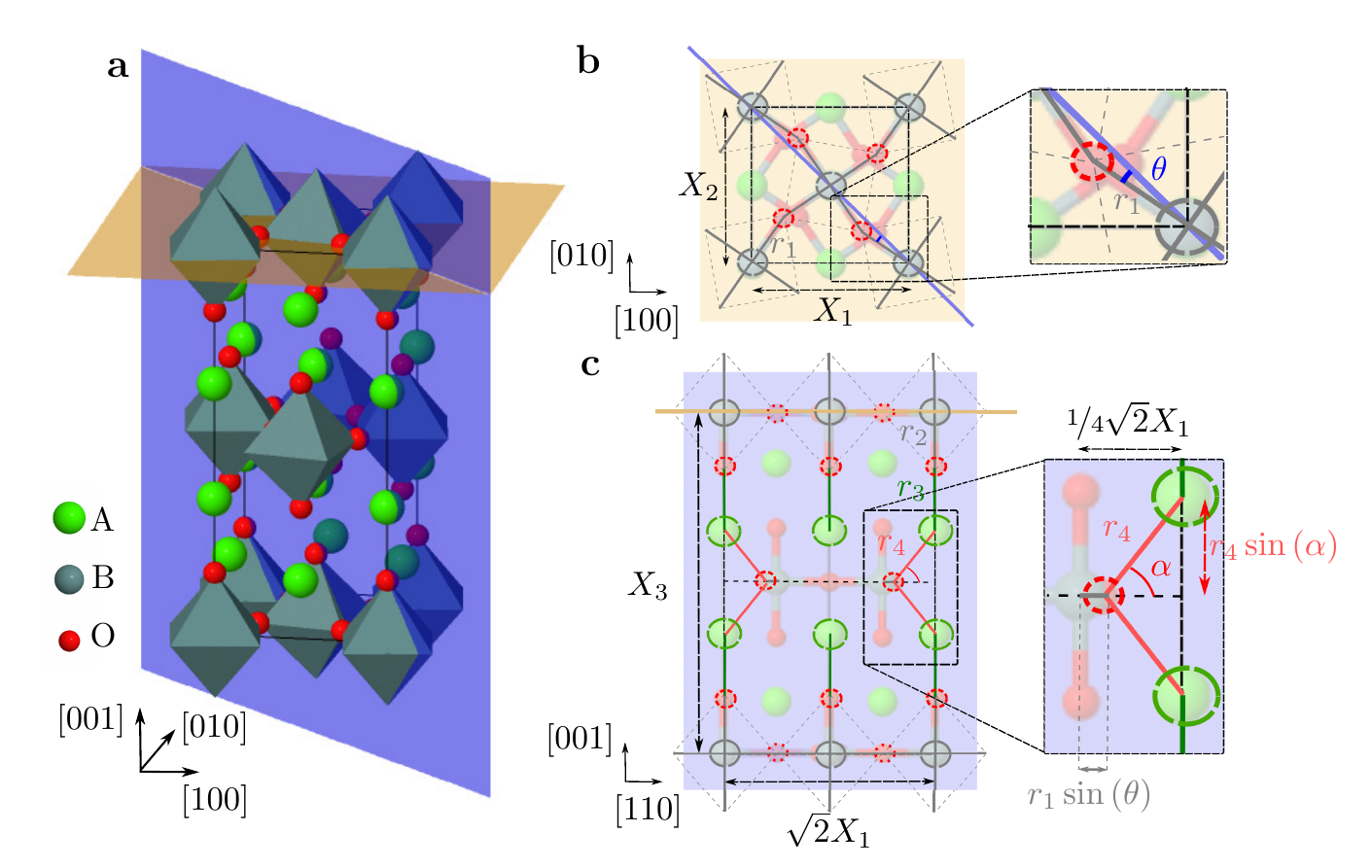}
\caption{{\footnotesize(a)} The unit cell of a A$_2$BO$_4$ Ruddlesden--Popper phase with a frozen rotation about the layering axis. A (green) and O (red) ions are shown as spheres and BO$_6$ octahedra are shown in grey. The $(001)$ (orange) and $(1\bar{1}0)$ (blue) planes are highlighted; {\footnotesize(b)} \& {\footnotesize(c)} Cross sections of the $(001)$ and $(1\bar{1}0)$ planes with insets magnifying sections of the illustrations. Atoms lying on the plane are shown as dashed circles and dashed squares show the outline of BO$_6$ octahedra. The lattice parameters $X_1$--$X_3$ and shortest bond lengths $r_1$--$r_4$ are labelled alongside the in-plane rotation angle, $\theta$ and the angle, $\alpha$, between the A cation and the $(002)$ plane (containing the B cation and equatorial O). Equatorial O ions in the $(001)$ plane in {\footnotesize(c)} have a more sparse dash since they really lie $r_1\sin(\theta)$ out of the $(1\bar{1}0)$ plane.}
\label{fig:diagram}
\end{figure}

Figure \ref{fig:diagram}a shows an $n=1$ A$_2$BO$_4$ Ruddlesden-Popper compound with a frozen octahedral rotation about the layering axis. This structure is visualised in the conventional manner for perovskite based chemistries: BO$_6$ octahedra are displayed as polyhedra (grey) with O ions at the corner linkages (red) and A cations (green) sitting in the interstitial sites with no A--O chemical bonds shown. The structure depicted in Figure \ref{fig:diagram} has an octahedral rotation that is in-phase between equivalent periodic image cells along $\left[001\right]$, placing the cell in the $Acam$ space-group symmetry. However, the mechanism described in this report would apply equally well to a phase in which the octahedral rotations were exactly out-of-phase between adjacent periodic cells. 

Figures \ref{fig:diagram}b and \ref{fig:diagram}c depict cross-sections of the $(001)$ and $(1\bar{1}0)$ planes that are highlighted in orange and blue respectively in Figure \ref{fig:diagram}a. Atoms lying on these planes are shown with thick dashed open circles and the shortest four metal-oxygen bonds that lie in these planes are shown as rigid struts. These four bonds consist of: the two distinct B--O bonds -- between the central B cation and the equatorial ($r_1$) and apical oxygen ($r_2$) corner ions; the shortest A--O distance -- across the AO layer interface between layers of adjacent BO$_6$ octahedra ($r_3$); and finally the shortest distance between the interface A cation and equatorial O anions ($r_4$). Equatorial O anions in the $(001)$ plane do not actually lie in $(1\bar{1}0)$ due to the non-zero value of the rotation angle $\theta$ and in fact sit a distance of $r_1 \sin \left(\theta\right)$ from this plane. These O anions are shown as thick dotted open circles and the struts of length $r_1$ connecting these equatorial O ions to nearest neighbour B cations are shown projected onto $(1\bar{1}0)$.

The four shortest metal cation bonds ($r_1$ -- $r_4$) were identified in our previous work to be the most stiff in DFT simulations on the $I4_1/acd$ phase of Ca$_2$GeO$_4$ \cite{Ablitt2017}. In this work, we assume that these four stiff bonds are perfectly rigid, leaving the only a single remaining internal degree of freedom in the structure: the in-plane octahedral rotation angle $\theta$. The other angle, $\alpha$, shown in the inset to Figure \ref{fig:diagram}c between the equatorial O plane and the shortest A--equatorial O bond ($r_4$), can be understood by trigonometry to be dependent on $\theta$. Changes in lattice parameters, $X_1$ and $X_3$, are therefore only facilitated through changes in this angle $\theta$, leading to a strong coupling between $X_1$ and $X_3$ via this internal degree of structural freedom. Following the method of previous works \cite{Smith2000, Grima2000, Grima2012}, we model changes in the $\widehat{B-O-B}$ and $\widehat{A-O-A}$ bond angles through harmonic potentials. That is to say that the work done per hinge, $w_{\varphi}$, is expressed in terms of incremental changes in bond angle $d \varphi$ by

\begin{equation}
w_{\varphi} = \text{\nicefrac{1}{2}} k_{\varphi} \left(d\varphi \right)^2
\label{eq:workperhinge}
\end{equation}

where $k_{\varphi}$ is the hinge stiffness. Since changes in these bond angles are the only allowed internal deformations, we are able to derive resulting mechanical properties for the crystal.

We may write an expression for the in-plane lattice parameter $X_1$ based on Figure \ref{fig:diagram}b,

\begin{equation}
X_1 = X_2 = 2\sqrt{2} r_1 \cos\left(\theta\right),
\label{eq:X1}
\end{equation}

since the unit cell is tetragonal, the two in-plane lattice parameters, $X_1$ and $X_2$, are equal.

A similar expression for $X_3$ can be formed by inspection of Figure \ref{fig:diagram}c

\begin{equation}
X_3 = 2 \left[r_2 + r_3 + r_4\sin\left(\alpha\right)\right]
\label{eq:X3}
\end{equation}

Further, equating the $[110]$ cell diagonal that constitutes the $x$-axis in Figure \ref{fig:diagram}c to the diagonal that the blue $(1\bar{1}0)$ plane cuts across Figure \ref{fig:diagram}b, the inset to Figure \ref{fig:diagram}c shows how one quarter of this length may be expressed in terms of both $\theta$ and $\alpha$,

\begin{equation}
\frac{\sqrt{2} X_1}{4} = r_4 \cos\left(\alpha\right) + r_1\sin\left(\theta\right).
\label{eq:alphasetup}
\end{equation}

Combining Equations \eqref{eq:X1} and \eqref{eq:alphasetup} leads to the relation

\begin{equation}
r_4 \cos\left(\alpha\right) = r_1 \left[\cos\left(\theta\right) - \sin\left(\theta\right) \right],
\label{eq:alphadef}
\end{equation}

and therefore from Equation \eqref{eq:alphadef} we may compute the derivative, $\frac{d\alpha}{d\theta}$,

\begin{equation}
\frac{d\alpha}{d\theta} = \left(\frac{r_1}{r_4}\right) \left( \frac{\sin\left(\theta\right) + \cos\left(\theta\right)}{\sin\left(\alpha\right)} \right).
\label{eq:dalphadtheta}
\end{equation}

The Poisson's ratio, relating the strain of lattice parameter $X_j$ to that of $X_i$ is defined as

\begin{equation}
\nu_{ij} = - \frac{d\varepsilon_j}{d\varepsilon_i},
\end{equation}

where the incremental strain of $X_i$, $d\varepsilon_i$, is defined in terms of the incremental extension $d X_i$

\begin{equation}
d\varepsilon_{i} = \frac{d X_i}{X_i}.
\label{eq:straindef}
\end{equation}

Substituting Equation \eqref{eq:X1} into Equation \eqref{eq:straindef}, we may thus define the incremental strains of the $X_1$ and $X_2$ lattice parameters as

\begin{equation}
d\varepsilon_{1} = d\varepsilon_2 = \frac{d X_1}{X_1} = \frac{1}{X_1} \frac{d X_1}{d \theta} d\theta =  \frac{- 2\sqrt{2} r_1 \sin\left(\theta\right)}{X_1} d\theta = - \tan\left(\theta\right) d\theta.
\label{eq:strain1}
\end{equation}

It is immediately apparent, since $d\varepsilon_{1} = d\varepsilon_2 $, that

\begin{equation}
\nu_{12} = \nu_{21} = -1.
\end{equation}

This is the same result that was derived by Grima and Evans \cite{Grima2000} for rigid squares on a 2D plane (recall Figure \ref{fig:squares}), which is exactly what our system reduces to on the $(001)$ cross-section shown in Figure \ref{fig:diagram}b. In the present three-dimensional scheme, however, by substituting Equations \eqref{eq:X3} and \eqref{eq:dalphadtheta} into Equation \eqref{eq:straindef}, we may also compute the incremental strain of the third lattice parameter, $X_3$,

\begin{equation}
\begin{split}
d\varepsilon_{3} = \frac{d X_3}{X_3} = \frac{1}{X_3} \left(\frac{d X_3}{d \alpha}\right) \left(\frac{d \alpha}{d \theta} \right) d\theta &= \frac{2 r_4 \cos\left(\alpha\right)}{X_3} \left(\frac{r_1}{r_4}\right) \left( \frac{\sin\left(\theta\right) + \cos\left(\theta\right)}{\sin\left(\alpha\right)} \right) d\theta \\
&= \frac{2 r_1}{X_3} \left( \frac{\sin\left(\theta\right) + \cos\left(\theta\right)}{\tan\left(\alpha\right)} \right) d\theta. \\
\end{split}
\label{eq:strain3}
\end{equation}

It is therefore possible to compute the Poisson ratio, $\nu_{13}$, relating $d\varepsilon_3$ to $d\varepsilon_1$,

\begin{equation}
\nu_{13} = - \frac{d\varepsilon_3}{d\varepsilon_1} = \left(\frac{X_1}{X_3}\right) \frac{2 r_1 \left( \frac{\sin\left(\theta\right) + \cos\left(\theta\right)}{\tan\left(\alpha\right)} \right) d\theta}{2\sqrt{2} r_1 \sin\left(\theta\right) d\theta} = \left(\frac{X_1}{X_3}\right) \left( \frac{1 + \cot\left(\theta\right)}{\sqrt{2} \tan\left(\alpha\right)} \right),
\label{eq:v13}
\end{equation}

and similarly for $\nu_{31}$, 

\begin{equation}
\nu_{31} = - \frac{d\varepsilon_1}{d\varepsilon_3} = \frac{1}{\nu_{13}} = \left(\frac{X_3}{X_1}\right) \left( \frac{\sqrt{2} \tan\left(\alpha\right)}{1 + \cot\left(\theta\right)} \right).
\label{eq:v31}
\end{equation}

In continuum elasticity, the strain energy, $U$, due to an incrementally small strain, $d\varepsilon_i$, is expressed in terms of the Young's modulus along $i$, $E_i$, as

\begin{equation}
U = \text{\nicefrac{1}{2}}E_i \left(d\varepsilon_i\right)^2 = \frac{W}{V}.
\label{eq:strainenergy}
\end{equation}

In Equation \eqref{eq:strainenergy}, using the principle of conservation of energy, $U$ has been equated to the work done by the cell, $W$, divided by the cell volume, $V$. In Equation \eqref{eq:workperhinge}, the work done per hinge, $w_{\varphi}$ is a quadratic function of the incremental change in hinge angle $d\varphi$ with stiffness $k_{\varphi}$, where $\varphi$ is the hinge angle. By inspection of Figure \ref{fig:diagram}b we see that the B--O--B $\theta$-hinge angle is $\varphi_{\theta} = 180 - 2\theta$, so that $\frac{d \varphi_{\theta}}{d \theta} = -2$, and therefore the work done by a $\theta$-hinge is

\begin{equation}
w_{\theta} = \frac{1}{2} k_{\theta} \left[\left(\frac{d \varphi_{\theta}}{d \theta}\right) d\theta \right]^2 =  2 k_{\theta} \left(d\theta\right)^2.
\end{equation}

Similarly, the A--O--A $\alpha$-hinge angle is $\varphi_{\alpha} = 2\alpha$, so that we may express $w_{\alpha}$ as

\begin{equation}
w_{\alpha} = 2 k_{\alpha} \left(d\alpha\right)^2.
\end{equation}

If we define $N_{\theta}$ and $N_{\alpha}$ as the number of $\theta$ and $\alpha$ hinges respectively, we can express W as

\begin{equation}
W = 2 \left[ N_{\theta} k_{\theta} \left(d\theta\right)^2 + N_{\alpha} k_{\alpha} \left(d\alpha\right)^2\right].
\label{eq:totalwork1}
\end{equation}

There are four $\theta$-hinges per layer and for the $n=1$ cell in Figure \ref{fig:diagram} there are two layers per unit cell, so $N_{\theta} = 8$. Similarly the $(1\bar{1}0)$ cross section in Figure \ref{fig:diagram}c shows two $\alpha$-hinges, both bisected by the $(002)$ plane (the middle BO$_6$ layer shown). The perpendicular cross section - the $(\bar{1}10)$ plane - also contains another two $\alpha$ hinges bisected by $(002)$. Furthermore, there should be the same four $\alpha$-hinges bisected by the $(001)$ plane (the top BO$_6$ layer coloured in yellow in Figure \ref{fig:diagram}c). By carefully studying Figure \ref{fig:diagram}a it should be apparent that these four $\alpha$-hinges bisected by $(001)$ lie in the $(220)$ and $(\bar{2}20)$ planes. Therefore $N_{\alpha} = N_{\theta} = 8$ and Equation \eqref{eq:totalwork1} may be rewritten

\begin{equation}
W = 16 \left[ k_{\theta}  + k_{\alpha} \left(\frac{d\alpha}{d\theta}\right)^2\right]  \left(d\theta\right)^2.
\label{eq:totalwork2}
\end{equation}

Substituting Equation \eqref{eq:totalwork2} into Equation \eqref{eq:strainenergy}, we may form the equation:

\begin{equation}
\begin{split}
\text{\nicefrac{1}{2}}~E_i \left(d\varepsilon_i\right)^2 &= \frac{16 \left[ k_{\theta}  + k_{\alpha} \left(\frac{d\alpha}{d\theta}\right)^2\right]  \left(d\theta\right)^2}{X_1^2 X_3}, \\
\end{split}
\end{equation}

and substituting for $d\varepsilon_1$ from Equation \eqref{eq:strain1},

\begin{equation}
\text{\nicefrac{1}{2}}~E_1 \left(\frac{8 r_1^2 \sin^2\left(\theta\right)\left(d\theta\right)^2}{X_1^2} \right) = \frac{16 \left[ k_{\theta}  + k_{\alpha} \left(\frac{d\alpha}{d\theta}\right)^2\right]  \left(d\theta\right)^2}{X_1^2 X_3},
\end{equation}

we may form an expression for $E_1$

\begin{equation}
E_1 = \frac{4 \left[ k_{\theta}  + k_{\alpha} \left(\frac{d\alpha}{d\theta}\right)^2\right] }{r_1^2 \sin^2\left(\theta\right) X_3}.
\label{eq:E1}
\end{equation}

Similarly, substituting for $d\varepsilon_3$ from Equation \eqref{eq:strain3}, we find an expression for $E_3$:

\begin{equation}
E_3 = \frac{8 \tan^2\left(\alpha\right) X_3 \left[ k_{\theta}  + k_{\alpha} \left(\frac{d\alpha}{d\theta}\right)^2\right] }{r_1^2 \left[\sin\left(\theta\right) + \cos\left(\theta\right) \right]^2 X_1^2}.
\label{eq:E3}
\end{equation}

The elastic compliance relates the strain experienced by a material to the applied stress. Expressing strain, $\varepsilon_i$, and stress, $\sigma_j$, as vectors in Voigt notation, we may define the elastic compliance matrix, $s_{ij}$, by the equation:

\begin{equation}
\varepsilon_i = s_{ij} \sigma_{j}
\end{equation}

The mechanical model depicted in Figure \ref{fig:diagram} does not allow shearing, meaning that the $\varepsilon_i$ ($i=4,5,6$) terms will always be $0$. Therefore we may restrict our strain/stress vectors to the first three terms in Voigt notation (the normal strains/stresses), making $s_{ij}$ a $3\times3$ matrix with all other components $0$. In terms of the Young's moduli and Poisson ratios, $s_{ij}$ may be expressed in general terms for an orthorhombic system with no strains:

\begin{equation}
\bm{s} = \left(\begin{array}{ccc}
\frac{1}{E_1} & \frac{-\nu_{21}}{E_2} & \frac{-\nu_{31}}{E_3} \\
\frac{-\nu_{12}}{E_1} & \frac{1}{E_2} & \frac{-\nu_{32}}{E_3} \\
\frac{-\nu_{13}}{E_1} & \frac{-\nu_{23}}{E_2} &  \frac{1}{E_3} \\
\end{array}\right)
\end{equation}

And thus for our current system, in terms of the Young's moduli and Poisson ratios already derived, this becomes:

\begin{equation}
\bm{s} = \left(\begin{array}{ccc}
\frac{1}{E_1} & \frac{1}{E_1} & \frac{-\nu_{31}}{E_3} \\
\frac{1}{E_1} & \frac{1}{E_1} & \frac{-\nu_{31}}{E_3} \\
\frac{-\nu_{13}}{E_1} & \frac{-\nu_{13}}{E_1} &  \frac{1}{E_3} \\
\end{array}\right)
\label{eq:emptys}
\end{equation}

Substituting the values of $\nu_{13}$, $\nu_{31}$, $E_1$ and $E_3$ from Equations \eqref{eq:v13}, \eqref{eq:v31}, \eqref{eq:E1} \& \eqref{eq:E3} into Equation \eqref{eq:emptys} gives the full compliance matrix. It is then possible to confirm that $\bm{s}$ satisfies the correct symmetry by verifying that:

\begin{equation}
\frac{\nu_{13}}{E_1} = \frac{\nu_{31}}{E_3}
\end{equation}

\section{Structural Limitations}

For given bond lengths $r_1$ -- $r_4$, the model has a single degree of structural freedom, $\theta$, from which all other structural parameters, such as $X_1$, $X_3$ and $\alpha$ may be computed. Although they are not restricted to be equal by symmetry, for simplicity of this analysis let us imagine that the two B--O bond lengths are equal ($r_1 = r_2$) and the two shortest A--O bond lengths are also equal ($r_3 = r_4$), giving only two independent bond lengths $r_1$ and $r_4$. Aside from limiting the parameter space, this assumption has little effect since in all equations in Section \ref{sec:Derivation}, $r_2$ and $r_3$ appear only as the combination $r_2 + r_3$ in the formula for $X_3$ in Equation \eqref{eq:X3}.

In this section we explore the structural limitations of the model with changing $\theta$ and \nicefrac{$r_4$}{$r_1$} assuming that all values of $r_1$ and $r_4$ are possible and that there are no interactions between any atoms not connected by a stiff rod ($r_1$ -- $r_4$ in Figure \ref{fig:diagram}).

\begin{figure}[h!]
\centering
\begin{subfigure}{\textwidth}
\centering
\caption{}
\label{fig:X1vstheta}
\includegraphics[width=0.8\textwidth]{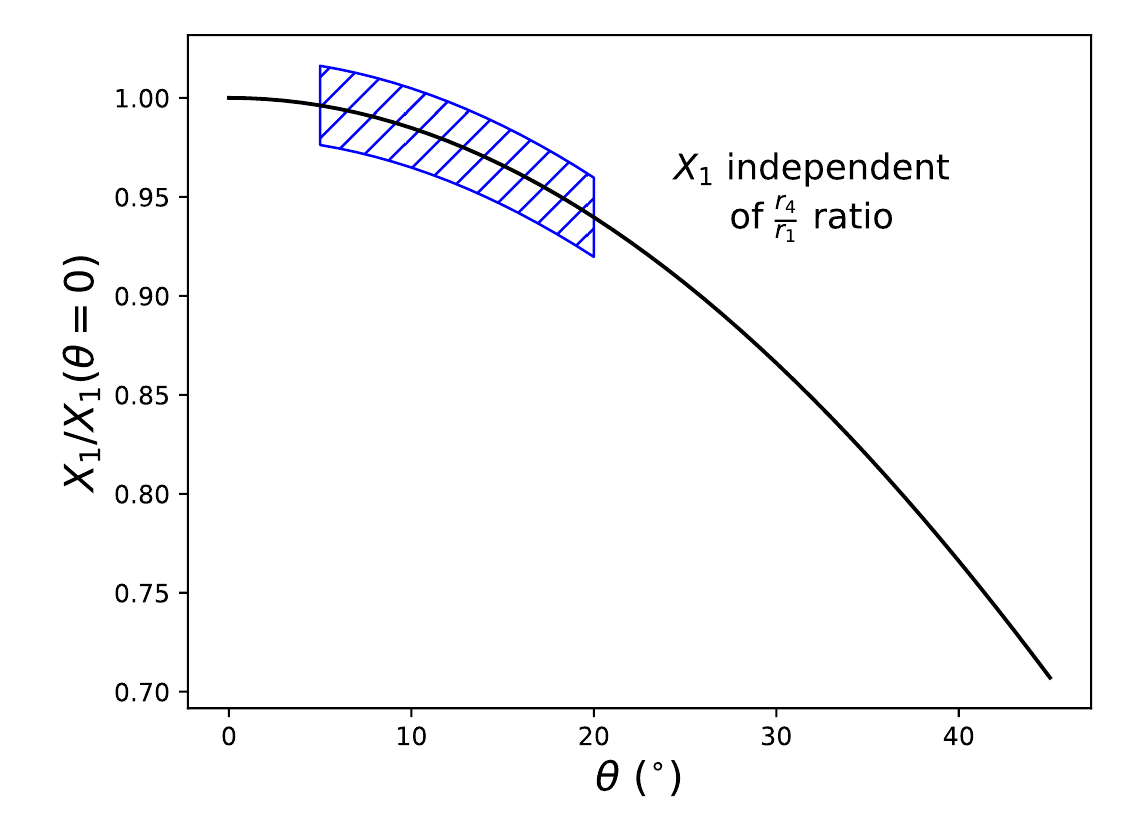}
\end{subfigure}\\[0.5cm]
\begin{subfigure}{\textwidth}
\centering
\caption{}
\label{fig:X1sketches}
\includegraphics[width=0.9\textwidth]{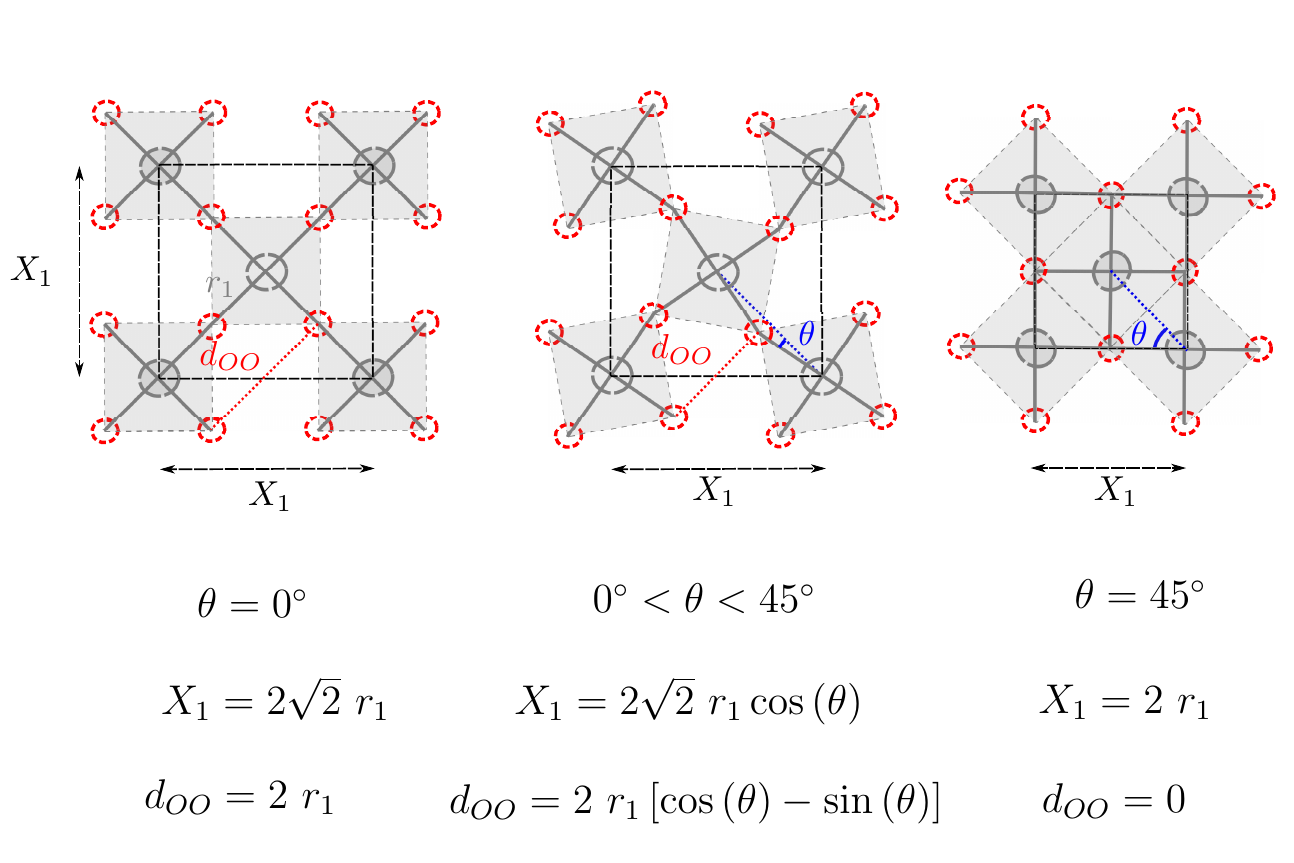}
\end{subfigure}\\[0.5cm]
\caption{{\footnotesize(a)} The in-plane lattice parameter, $X_1$, with changing $\theta$ in terms of $X_1$ when $\theta = 0^{\circ}$; {\footnotesize(b)} sketches of the $(001)$-plane cross section for different regimes of $\theta$. The blue hashed region in {\footnotesize(a)} shows the range of $\theta$ where the mechanism may reasonably operate in a real system.}
\label{fig:X1Analysis}
\end{figure}

Figure \ref{fig:X1vstheta} shows how the in-plane lattice parameter, $X_1$, varies with $\theta$ between the extreme values of $\theta = 0^{\circ}$ (corresponding to an unrotated, high-symmetry parent phase) and $\theta = 45^{\circ}$ (where the square cross sections of the BO$_6$ octahedra in the (001) plane are perfectly packed). The atoms lying on the (001) plane, the same plane shown previously in Figure \ref{fig:illustration}b, are illustrated for these extreme values of $\theta$ in Figure \ref{fig:X1sketches} alongside a structure with an intermediate $\theta$ value. $X_1$ is independent of the \nicefrac{$r_4$}{$r_1$} ratio and may decrease to $\frac{1}{\sqrt{2}}$ of its value in the unrotated structure by increasing $\theta$.

\begin{figure}[h!]
\centering
\begin{subfigure}{0.48\textwidth}
\centering
\includegraphics[width=0.98\textwidth]{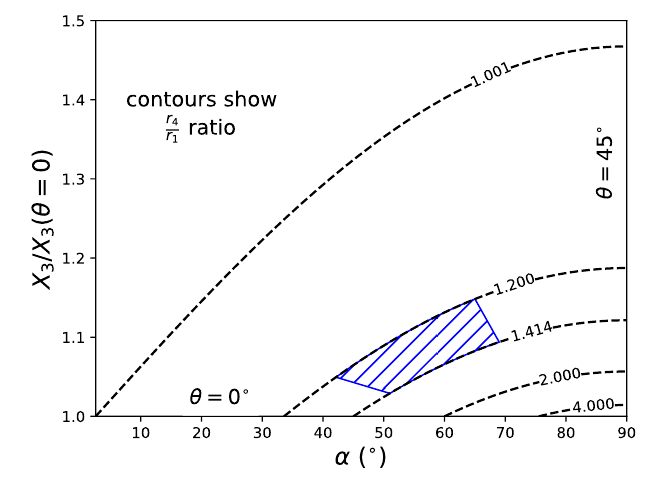}
\caption{}
\label{fig:X3vsalpha}
\end{subfigure}
\begin{subfigure}{0.48\textwidth}
\centering
\includegraphics[width=0.98\textwidth]{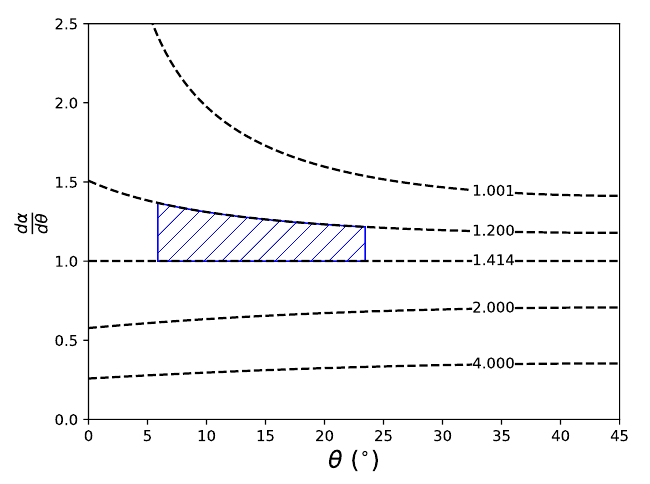}
\caption{}
\label{fig:dalphadthetavstheta}
\end{subfigure}\\[0.5cm]
\begin{subfigure}{\textwidth}
\centering
\includegraphics[width=0.95\textwidth]{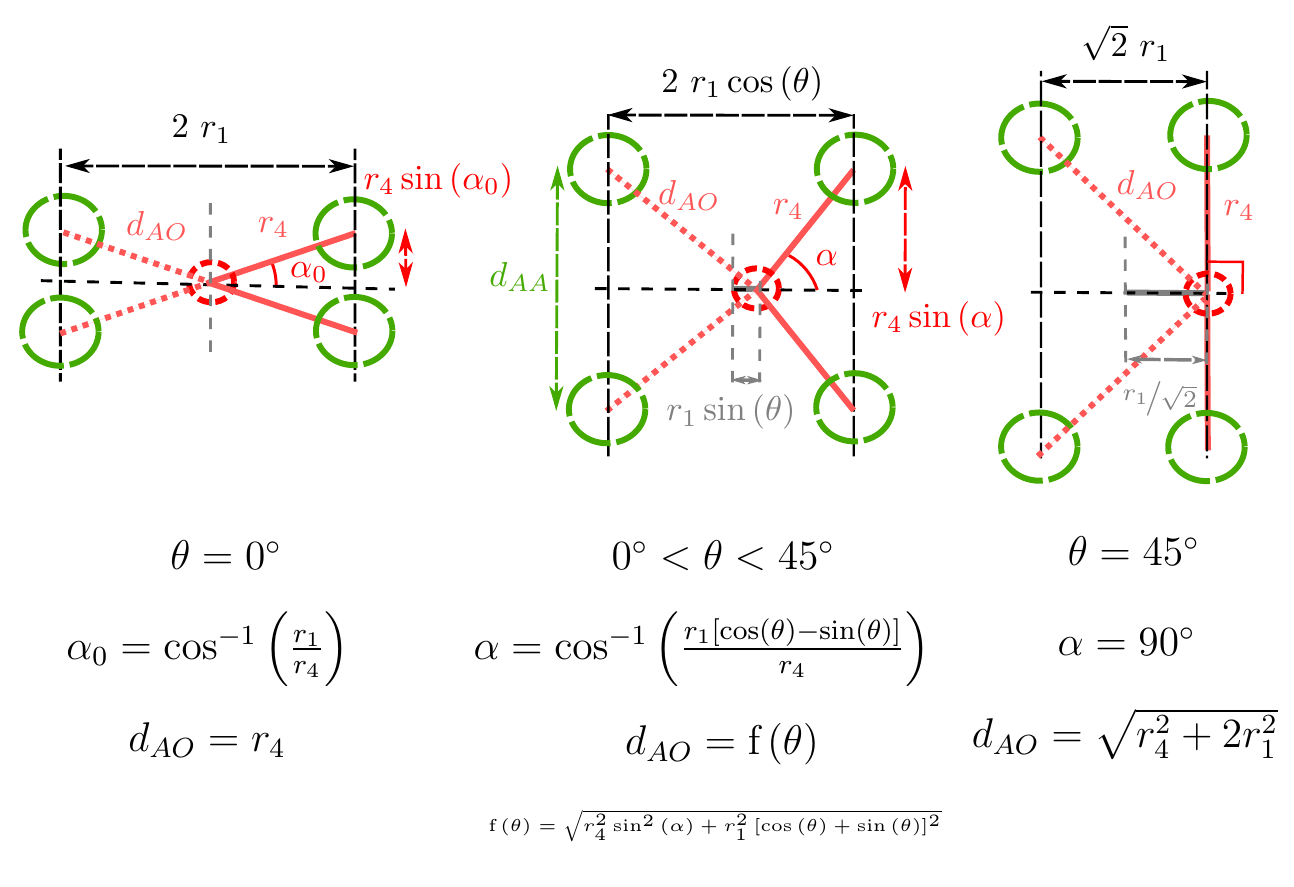}
\caption{}
\label{fig:X3sketches}
\end{subfigure}\\[0.5cm]
\caption{{\footnotesize(a)} The lattice parameter along the layering axis, $X_3$, with changing $\alpha$ and \nicefrac{$r_4$}{$r_1$} as $\theta$ is varied between $0^{\circ}$ and $45^{\circ}$; {\footnotesize(b)} $\frac{d\alpha}{d\theta}$ as a function of $\theta$ for different \nicefrac{$r_4$}{$r_1$} ratios; {\footnotesize(c)} sketches of the $(1\bar{1}0)$-plane cross section for different regimes of $\theta$. The blue hashed regions in {\footnotesize(a)} \& {\footnotesize(b)} show the ranges of $\theta$ and \nicefrac{$r_4$}{$r_1$} where the mechanism may reasonably operate in a real system.}
\label{fig:X3Analysis}
\end{figure}

Figure \ref{fig:X3vsalpha} then shows how the lattice parameter along the layering axis, $X_3$, varies between $\theta = 0^{\circ}$ and $\theta = 45^{\circ}$. Again Figure \ref{fig:X3sketches} shows a subset the atoms lying on the $(1\bar{1}0)$ plane, where this subset is an extension of the inset of Figure \ref{fig:illustration}c, for $\theta = 0^{\circ}, 45^{\circ}$ and an intermediate value. Since $X_3$ depends upon $\alpha$, which is itself a function of $\theta$ and \nicefrac{$r_4$}{$r_1$}, contours in Figure \ref{fig:X3vsalpha} show different \nicefrac{$r_4$}{$r_1$} ratios.

It should be apparent from the $\theta = 0^{\circ}$ panel of Figure \ref{fig:X3sketches} that $r_4 \geq r_1$. The $\theta = 0^{\circ}$ value of $\alpha$, $\alpha_0$, then depends on \nicefrac{$r_4$}{$r_1$}, where $\alpha = 0^{\circ}$ only if $\theta = 0^{\circ}$ and $r_4 = r_1$. In the $\theta = 45^{\circ}$ limit, $\alpha = 90^{\circ}$ always as the A--O ($r_4$) bond aligns with the $[001]$ direction. The greatest change in $X_3$ over the range of $\theta$ occurs when $r_4 = r_1$ such that $X_3$ may extend to $1.5 \,X_3{\left(\theta = 0^{\circ}\right)}$, although the rate of change in $X_3$ with $\alpha$ is greatest for low $\alpha$, where $\sin\left(\alpha\right)$ is approximately linear. The rate of change of $X_3$ with $\theta$ is even greater for low $\theta$ and low \nicefrac{$r_4$}{$r_1$} since Figure \ref{fig:dalphadthetavstheta} shows that $\frac{d\alpha}{d\theta} >> 1$. In contrast, a large \nicefrac{$r_4$}{$r_1$} ratio leads to a large $\alpha_0$, decreasing both the range of $\alpha$ as $\theta$ varies $0^{\circ} \rightarrow 45^{\circ}$ and the rate of change of $X_3$ with $\alpha$ at higher $\alpha$ values.

\section{Physical Limitations}
\label{sec:PhysicalLims}

The previous section gave the maximum changes in $X_1$ and $X_3$ that may be achieved from varying $\theta$ over the range $[0^{\circ}, 45^{\circ}]$ and found that the greatest proportional change in $X_3$ is achieved when $r_4 = r_1$. However, these limitations were extracted from the equations derived in Section \ref{sec:Derivation} and did not consider whether the model could still be valid across the full range of $\theta$ and \nicefrac{$r_4$}{$r_1$}.

\begin{figure}[h!]
\centering
\includegraphics[width=0.8\textwidth]{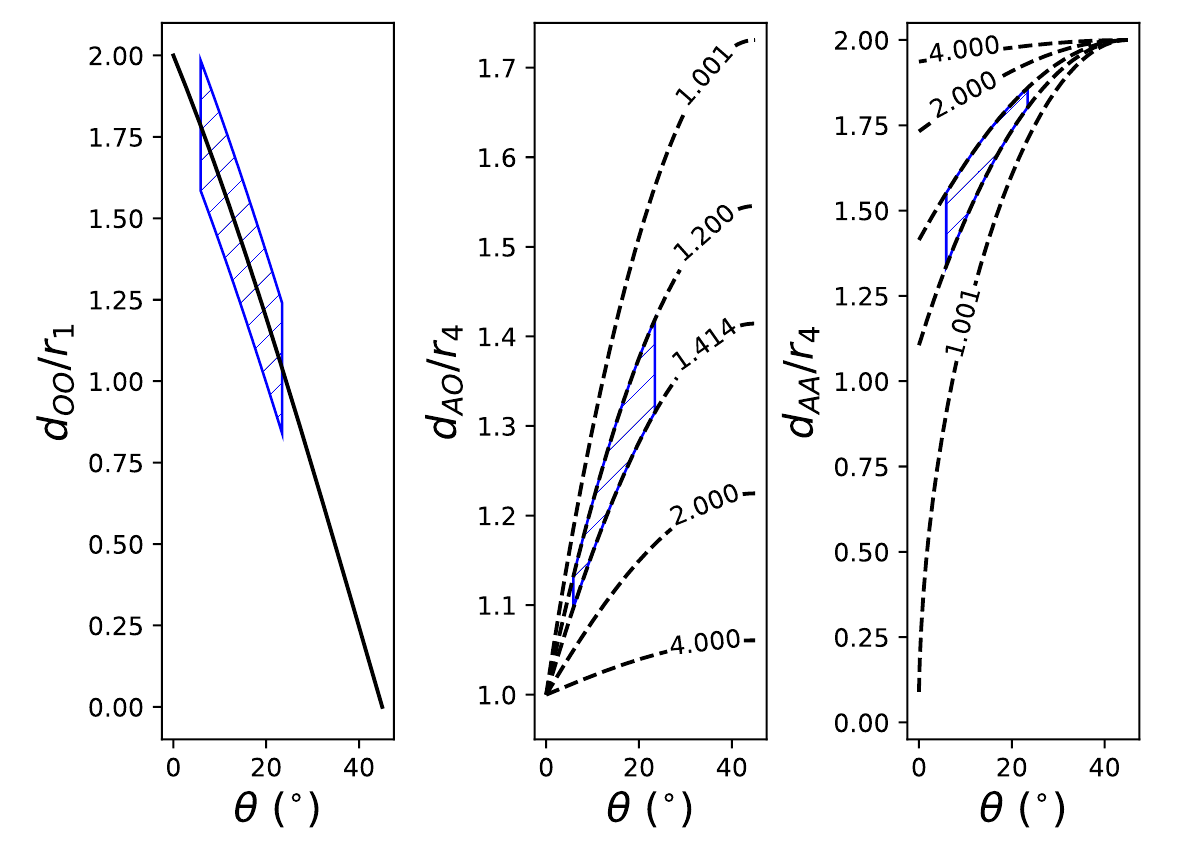}
\caption{Interatomic distances $d_{OO}$, $d_{AO}$ and $d_{AA}$, expressed in terms of fixed bond lengths $r_1$ or $r_4$ as a function of $\theta$ and \nicefrac{$r_4$}{$r_1$}. None of $d_{OO}$, $d_{AO}$ and $d_{AA}$ are restricted in the model, however the assumptions of the model become unreasonable if $d_{OO}$ or $d_{AA}$ become too small or if $d_{AO} \approx r_4$. Based on these criteria, the blue hashed region shows the ranges of $\theta$ and \nicefrac{$r_4$}{$r_1$} where the mechanism may reasonably operate in a real system.}
\label{fig:dOOdAAvstheta}
\end{figure}

Figure \ref{fig:X1sketches} shows the shortest distance between equatorial O ions on neighbouring BO$_6$ octahedra, labelled $d_{OO}$. The corkscrew model does not consider these two O ions to interact. If $\theta = 0^{\circ}$, $d_{OO} = 2 r_1$. However as $\theta$ increases, it can be seen from Figure \ref{fig:dOOdAAvstheta} that $d_{OO}$ becomes smaller until, in the limit $\theta = 45^{\circ}$, the two O ions lie on top of one another and thus $d_{OO} = 0$. This limit is clearly unphysical and one would expect that the two O ions should strongly repel one another at short distances. Therefore the assumption that the O ions have a negligible interaction is invalid at high $\theta$. $r_1$ is the shortest \emph{stiff} chemical bond in the system and even restricting that $d_{OO} > r_1$ imposes an upper limit of $\theta < 24^{\circ}$.

For a general value of $\theta$, there are two inequivalent distances between A and equatorial Os. The shorter of these two distances is labelled $r_4$ in Figure \ref{fig:X3sketches} and the longer is labelled $d_{AO}$. In the limit that $\theta = 0^{\circ}$, $d_{AO} = r_4$ for all \nicefrac{$r_4$}{$r_1$}, and $d_{AO}$ becomes greater than $r_4$ as $\theta$ increases. The corkscrew model assumes that the shorter bond is very stiff and thus $r_4$ remains constant, whereas the longer bond is so much longer that it has negligible stiffness and $d_{AO}$ is unrestricted. This assumption breaks down at very low $\theta$ where $d_{AO} \approx r_4$. For a given $\theta$, Figure \ref{fig:dOOdAAvstheta} shows that \nicefrac{$d_{AO}$}{$r_4$} is greatest when $r_4 \approx r_1$, but as \nicefrac{$r_4$}{$r_1$} becomes greater the maximum $d_{AO}$ differs less from $r_4$. Although it is difficult to estimate the stiffness of A--O bonds for different $d_{AO}$ distances without accurate electronic structure calculations, and therefore difficult to give a lower limit on the reasonable \nicefrac{$d_{AO}$}{$r_4$} ratio at which the model is valid, it is clear that the need for $d_{AO} >> r_4$ places a lower limit on the range of $\theta$ and an upper limit on the range of \nicefrac{$r_4$}{$r_1$}.

Finally, in Figure \ref{fig:X3sketches}, the distance between A cations above and below the (001) plane containing the equatorial Os is labelled as $d_{AA}$. Like $d_{OO}$ discussed previously, it is unphysical for these A cations to become arbitrarily close to one another. Figure \ref{fig:dOOdAAvstheta} shows that in the lower limit of \nicefrac{$r_4$}{$r_1$}, $d_{AA}$ becomes very small for low values of $\theta$ such that $d_{AA}\left(\theta=0^{\circ}\right) = 0$ if $r_4 = r_1$. If we assume that an unrotated parent phase can exist, even if the corkscrew model does not apply in this parent phase, low values of $\theta$ must be structurally achievable. Consequently, there is a lower limit on the possible \nicefrac{$r_4$}{$r_1$}.

\begin{figure}[h!]
\centering

\begin{subfigure}{\textwidth}
\centering
\caption{}
\label{fig:s11ands33vstheta}
\includegraphics[width=0.8\textwidth]{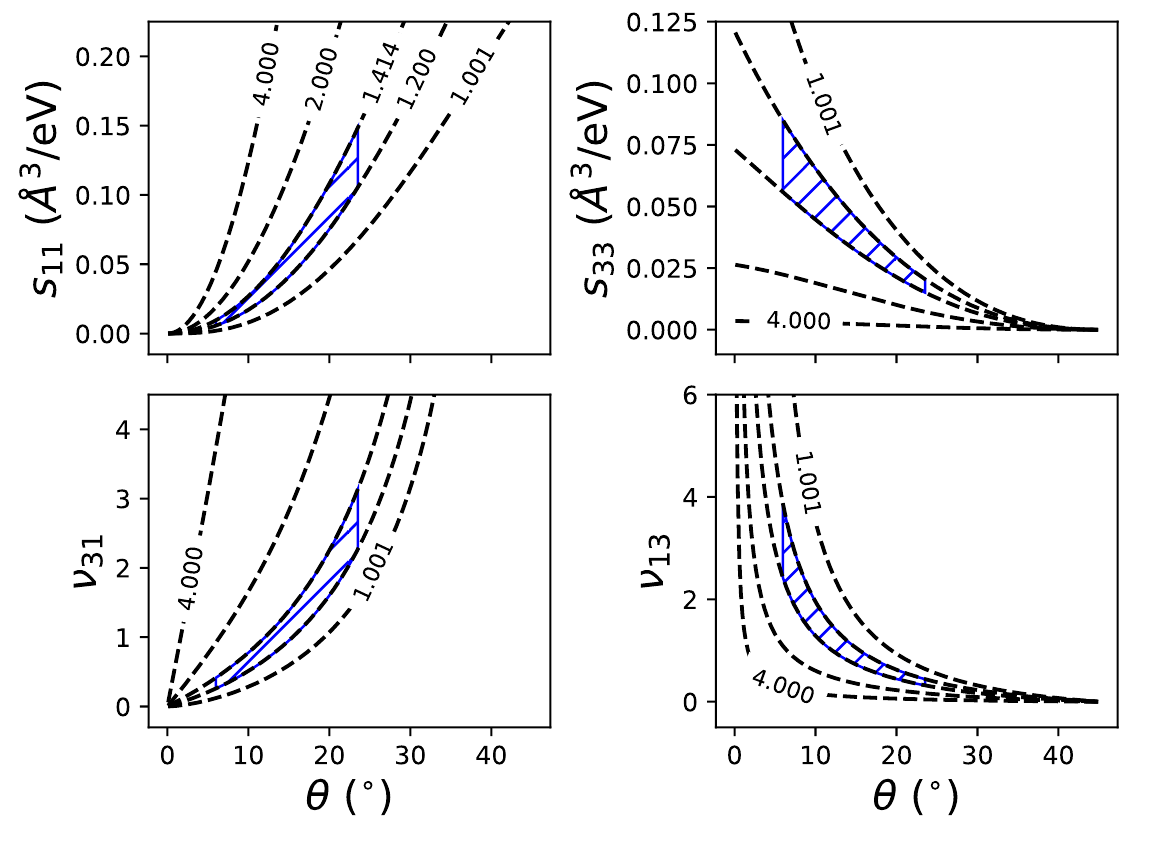}
\end{subfigure}\\[0.5cm]

\begin{subfigure}{\textwidth}
\centering
\caption{}
\label{fig:s13vstheta}
\includegraphics[width=0.8\textwidth]{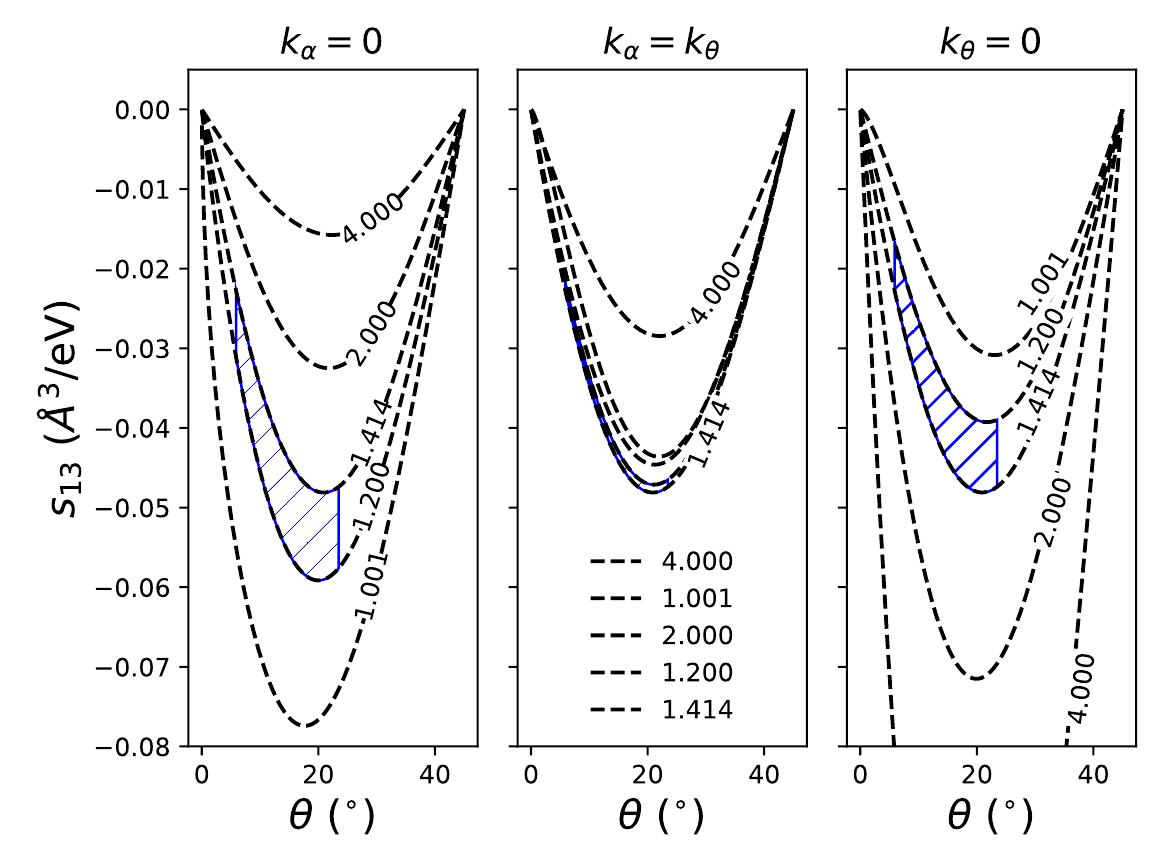}
\end{subfigure}

\caption{{\footnotesize(a)} $s_{11}$ \& $s_{33}$ diagonal components of the compliance matrix and $\nu_{31}$ \& $\nu_{31}$ Poisson's ratios and  {\footnotesize(b)} $s_{13}$ off-diagonal coupling component of the compliance matrix, all as a function of $\theta$ for different \nicefrac{$r_4$}{$r_1$} ratios. The blue hashed regions in {\footnotesize(a)} \& {\footnotesize(b)} show the ranges of $\theta$ and \nicefrac{$r_4$}{$r_1$} where the mechanism may reasonably operate in a real system. In all panels in {\footnotesize(a)} $k_{\theta} = k_{\alpha} = 1$ whereas in {\footnotesize(b)} different panels show different \nicefrac{$k_{\alpha}$}{$k_{\theta}$} ratios, always under the constraint that $k_{\theta} + k_{\alpha} = 2$.}
\label{fig:compliancesvstheta}
\end{figure}

The requirements that $d_{AO} >> r_4$ for $\theta$ at which the model applies and $d_{AA} >> 0$ for all $\theta$ place restrictions on the lower and upper values of \nicefrac{$r_4$}{$r_1$}. Since the Ruddlesden-Popper phases being discussed are layered perovskites with frozen octahedral rotations, a realistic range of \nicefrac{$r_4$}{$r_1$} could be estimated from typical values of Goldschmidt tolerance factors, $t$. In perovskites with frozen octahedral rotations, $t < 1.0$ and perovskites are rarely stable with tolerance factors below $t \approx 0.85$. These restrict \nicefrac{$r_4$}{$r_1$} to the range $1.2 \leq$ \nicefrac{$r_4$}{$r_1$} $\leq \sqrt{2}$. We have already placed an upper limit $\theta < 24^{\circ}$ to prevent $d_{OO}$ becoming too small. If $r_4 = \sqrt{2}\,r_1$, restricting $\theta > 6^{\circ}$ ensures that $d_{AO}$ is at least 10\% larger than $r_4$. These limits are only guesses at the values of $\theta$ and \nicefrac{$r_4$}{$r_1$} for which one expects the assumptions of the model to be violated in a real ionic material; blue hashed regions have been added to all graphs to indicate ballpark values accessible by a physically \emph{plausible} system.

\section{Prediction of Compliances}

In order to compute compliance components, we fix that $r_1 = 1.0$ and the hinge stiffnesses in Equation \eqref{eq:totalwork2} are also assigned arbitrary values of $k_{\theta} = k_{\alpha} = 1.0$.

Figure \ref{fig:s11ands33vstheta} presents the behaviour of the $s_{11}$ and $s_{33}$ compliance components as a function of $\theta$ for different values of \nicefrac{$r_4$}{$r_1$}. In the limit $\theta \rightarrow 0^{\circ}$, $s_{11} \rightarrow 0$ for all \nicefrac{$r_4$}{$r_1$}, yet $s_{11}$ increases rapidly as $\theta$ becomes larger, with the highest \nicefrac{$r_4$}{$r_1$} corresponding to the greatest $s_{11}$. In contrast $s_{33}$ shows the opposite behaviour as $s_{33} \rightarrow 0$ since $\theta \rightarrow 45^{\circ}$ for all \nicefrac{$r_4$}{$r_1$} and $s_{33}$ is very large for low $\theta$ and low \nicefrac{$r_4$}{$r_1$}; although for larger \nicefrac{$r_4$}{$r_1$} the change in $s_{33}$ with $\theta$ is very small.

Despite the appearance of Figure \ref{fig:s11ands33vstheta}, $s_{11}$ is finite for all values of $\theta$ and \nicefrac{$r_4$}{$r_1$}. This should be evident from Equation \eqref{eq:E1}, since $s_{11} =$ \nicefrac{1}{$E_1$} and $E_1$ is always non-zero providing that $k_{\theta}, k_{\alpha} > 0$. Similarly, $s_{33}$ is finite except in the limit $r_4 = r_1$ and $\theta \rightarrow 0^{\circ}$, since this is the only limit in which $E_3$ is zero. The limit $r_4 = r_1$, as well as the $\theta = 0^{\circ}$ and $\theta = 45^{\circ}$ extremes where the corkscrew model is perfectly stiff along one axis, were ruled out for a physical system in Section \ref{sec:PhysicalLims} and the blue hashed regions in Figure \ref{fig:s11ands33vstheta} correspond to finite and non-zero compliances. However, for both $s_{11}$ and $s_{33}$, within this hashed region, the compliance is very sensitive to changes in $\theta$. The lower two panels of Figure \ref{fig:s11ands33vstheta} show the Poisson ratios $\nu_{13}$ and $\nu_{31}$. These are both positive for all $\theta$ and \nicefrac{$r_4$}{$r_1$} and thus this mechanism does not lead to auxetic behaviour coupling in-plane and layering axes. Inspecting Equations \eqref{eq:v13} and \eqref{eq:v31} we see that both $\nu_{13}$ and $\nu_{31}$ are undefined for all \nicefrac{$r_4$}{$r_1$} in the limits $\theta \rightarrow 0^{\circ}$ and $\theta \rightarrow 45^{\circ}$ respectively. 

The three panels of Figure \ref{fig:s13vstheta} show the off-diagonal compliance matrix component, $s_{13}$, against $\theta$ and \nicefrac{$r_4$}{$r_1$} for the cases that $(k_{\theta}, k_{\alpha})$ = $(2, 0)$, $(1,1)$ and $(0,2)$. Whereas the $s_{11}$ ($s_{33}$) axial compliance increases (decreases) monotonically with $\theta$, in both the limits $\theta = 0^{\circ}$ and $ \theta = 45^{\circ}$ $s_{13} = 0$ for all \nicefrac{$r_4$}{$r_1$}, yet decreases to reach a minimum between these limits. From Equation \ref{eq:emptys}, we see that $s_{13} = -\nu_{13} \, s_{11} = -\nu_{31} \, s_{33}$. Hence, $\nu_{ij} \rightarrow \infty$ in the same limit that the corresponding axial compliance $s_{ii} \rightarrow 0$ meaning that $s_{13}$ has moderate values for all $\theta$ and \nicefrac{$r_4$}{$r_1$} and that $s_{13} \rightarrow 0$ at the limits of $\theta$.

So far we have not discussed the effect of $k_{\theta}$ and $k_{\alpha}$ on the elastic compliances. In the case that $k_{\alpha} = 0$ (the first panel in Figure \ref{fig:s13vstheta}), the work, $W$, in Equation \eqref{eq:totalwork2} is constant for all $\theta$ and \nicefrac{$r_4$}{$r_1$}. In this case $|s_{13}|$ is maximised when $r_4 \approx r_1$, however the elastic coupling between $X_1$ and $X_3$ becomes weaker as $r_4 >> r_1$. Since $|s_{13}|$ is greatest for $\theta \approx 20^{\circ}$, inspecting the blue hashed region of Figure \ref{fig:s13vstheta} we see that the strength of compliance coupling is greatest for the largest $\theta$ within the range of potentially physical values identified previously.

If $k_{\alpha} > 0$, $W$ contains a $\frac{d \alpha}{d \theta}$ term that was plotted as a function of $\theta$ and \nicefrac{$r_4$}{$r_1$} in Figure \ref{fig:dalphadthetavstheta}. In the special case that $r_4 = \sqrt{2} \, r_1$, $\frac{d \alpha}{d \theta} = 1$ for all $\theta$ and therefore, since $k_{\theta} + k_{\alpha} = 2$ in every panel, $W$ is constant and the $r_4 = \sqrt{2} \,r_1$ contour is unchanged between panels in Figure \ref{fig:s13vstheta}. For contours where $r_4 \neq \sqrt{2} \,r_1$, increasing the $k_{\alpha}$ weighting significantly affects the $s_{13}$ dependence upon \nicefrac{$r_4$}{$r_1$} such that in the final $(k_{\theta}, k_{\alpha})$ = $(0,2)$ panel, the order of contours is flipped as compared to the first $(2, 0)$ graph. The intermediate, $(1,1)$ case, also used to compute the axial compliances in Figure \ref{fig:s11ands33vstheta}, has a change in ordering of the \nicefrac{$r_4$}{$r_1$} contours yet again, with the $r_4 = \sqrt{2}\, r_1$ contour now corresponding to the highest $|s_{13}|$ values. Since the blue hashed region is bounded by the invariant $r_4 = \sqrt{2} \, r_1$ contour, the compliance components of systems with parameters in this region are only weakly dependent upon \nicefrac{$k_{\alpha}$}{$k_{\theta}$}.

\newpage

\section{Extending to arbitrary $n$}

\begin{wrapfigure}{l]}{0.5\textwidth}
\begin{center}
\includegraphics[height=8cm]{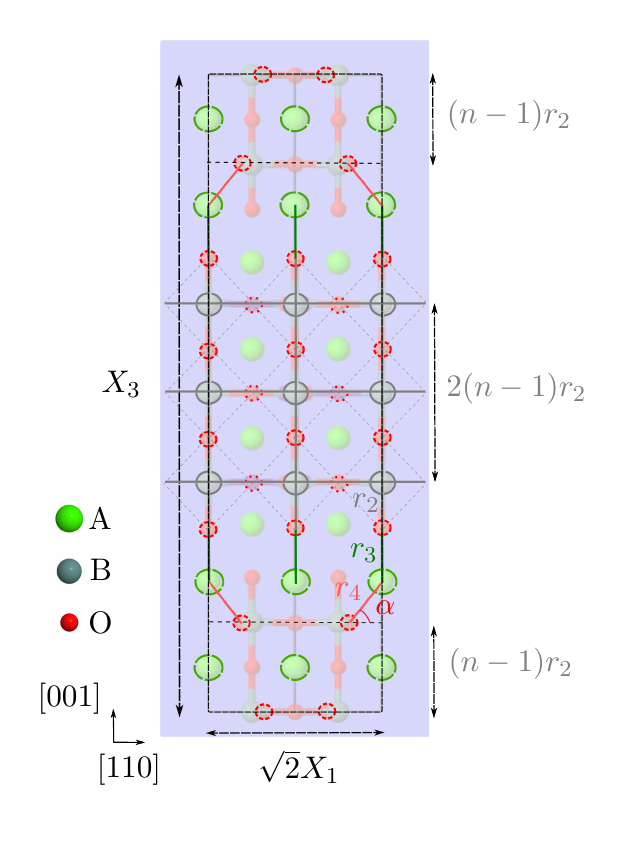}
\end{center}
\caption{The cross section of the $(1\bar{1}0)$ plane of an $n=3$ phase with a frozen rotation about the layering axis to represent a general member of the A$_{n+1}$B$_n$O$_{3n+1}$ series. Atoms lying on the plane are shown as dashed circles and dashed squares show the outline of BO$_6$ octahedra. The lattice parameters $X_1$--$X_3$ and shortest bond lengths $r_1$--$r_4$ are labelled alongside the in-plane rotation angle, $\theta$ and the angle, $\alpha$, between the A cation and the plane containing the B cation and equatorial O.}
\label{fig:Corkscrew3}
\end{wrapfigure}

The derivation in Section \ref{sec:Derivation} focused only on the $n=1$ member of the A$_{n+1}$B$_n$O$_{3n+1}$ series, however it is possible to extend this derivation to a system of general layer thickness $n$. Figure \ref{fig:Corkscrew3} shows the $(\bar{1}10)$ plane of the $Acam$ phase of $n=3$ A$_{4}$B$_3$O$_{10}$, the same plane displayed in Figure \ref{fig:diagram}c for $n=1$ A$_{2}$BO$_{4}$. 

The in-plane structure (i.e. the cross section through the $(001)$ and $(002)$ planes shown in Figure \ref{fig:diagram}b is exactly the same in the $n=3$ and $n=1$ cases. Therefore for $n=3$, $X_1$ and $X_2$ will be given by Equation \eqref{eq:X1} and hence $\nu_{12} = -1$ also. 

Additional planes of BO$_6$ octahedra clearly affect the expression for $X_3$, however. Noting that the thickness of the cell edge and middle perovskite layer blocks have both increased by $n-1$ $BO_6$ units (each $2r_2$ in height) between Figure \ref{fig:diagram}c and Figure \ref{fig:Corkscrew3}, but the same number of $r_3$ and $r_4$ bonds remain. We may express $X_3$ generally as:

\begin{equation}
X_3 = 2 \left[\left(2n - 1\right)r_2 + r_3 + r_4\sin\left(\alpha\right)\right]
\label{eq:X3n}
\end{equation}

Similarly, since each $BO_6$ layer contains 4 $\theta$-hinges, there are $n$ $BO_6$ layers per perovskite block and 2 perovskite blocks in the unit cell, Equation \eqref{eq:totalwork2} may be generalised:

\begin{equation}
W = 16 \left[ n\,k_{\theta}  + k_{\alpha} \left(\frac{d\alpha}{d\theta}\right)^2\right]  \left(d\theta\right)^2
\label{eq:Wn}
\end{equation}

It is then straightforward to use the new definitions of $X_3$ and $W$ to construct the terms of the $\bm{s}$ matrix in Equation \eqref{eq:emptys} as a function of any $n$.

\begin{equation}
s_{11} = s_{12} = \left( \frac{ \sin^2\left(\theta\right)}{4} \right) ~ \frac{r_1^2 ~ X_3}{\left[ n\,k_{\theta}  + k_{\alpha} \left(\frac{d\alpha}{d\theta}\right)^2\right]},
\label{eq:E1n}
\end{equation}

\begin{equation}
s_{33} = \left( \frac{\cos^2\left(\theta\right) \left[\sin\left(\theta\right) + \cos\left(\theta\right) \right]^2}{\tan^2\left(\alpha\right) } \right) ~ \frac{r_1^4}{X_3 \left[ n\,k_{\theta}  + k_{\alpha} \left(\frac{d\alpha}{d\theta}\right)^2\right] },
\label{eq:E3n}
\end{equation}

\begin{equation}
s_{13} = - \left( \frac{\sin\left(2\theta\right)\left[\sin\left(\theta\right) + \cos\left(\theta\right) \right]}{4 \tan\left(\alpha\right)} \right) ~ \frac{r_1^3}{\left[n\,k_{\theta}  + k_{\alpha} \left(\frac{d\alpha}{d\theta}\right)^2 \right]}.
\end{equation}

Inspecting Equation \eqref{eq:X3n}, we see that it has the form $X_3 = a\,n + b$ where $b$ contains the component to $X_3$ that may change with changing $\theta$ and that this term forms a smaller proportion of $X_3$ as $n$ increases. It was shown previously that the behaviour of $W$ with changing $\theta$ is weakly dependent upon \nicefrac{$k_{\alpha}$}{$k_{\theta}$} when \nicefrac{$r_4$}{$r_1$} $\neq \sqrt{2}$. On the other hand, Equation \eqref{eq:Wn} shows that when comparing structures with different $n$, the \nicefrac{$k_{\alpha}$}{$k_{\theta}$} ratio is extremely important such that if \nicefrac{$k_{\alpha}$}{$k_{\theta}$} $<< 1$, $W \propto n$ whereas if \nicefrac{$k_{\alpha}$}{$k_{\theta}$} $>> 1$, $W \propto 1$. Using this insight, in Table \ref{tab:sndependence} we show the dependence of $s_{11}$, $s_{33}$ and $s_{13}$ on $n$ for different \nicefrac{$k_{\alpha}$}{$k_{\theta}$} regimes.

\begin{table}
\centering
\begin{tabular}{cccc}
\nicefrac{$k_{\alpha}$}{$k_{\theta}$} $<< 1$ & $s_{11} \propto 1 + \frac{c}{n}$ &  $s_{33} \propto \frac{1}{n\left(n+c\right)}$ & $s_{13} \propto -\frac{1}{n}$ \\[0.2cm]
\nicefrac{$k_{\alpha}$}{$k_{\theta}$} $>> 1$ & $s_{11} \propto n + c$ &  $s_{33} \propto \frac{1}{n + c}$ & $s_{13} \propto 1$ \\
\end{tabular}
\caption{Proportionality of various components of the elastic compliance matrix $s_{ij}$ to the perovskite layer thickness $n$ for different regimes in the ratio of harmonic hinge stiffnesses related to the $\alpha$ and $\theta$ bond angles.}
\label{tab:sndependence}
\end{table}

Since $n \geq 1$, there is no problem if any of the compliances cease to make sense in the limit $n \rightarrow 0$. However, pure ABO$_3$ perovskite represents the $n = \infty$ limit of the RP series and therefore the compliances in this limit should have a physical interpretation. The only term in Table \ref{tab:sndependence} that looks problematic as $n \rightarrow \infty$ is $s_{11} \propto n + c$ under the condition \nicefrac{$k_{\alpha}$}{$k_{\theta}$} $>> 1$. It should be noted in this case that providing $k_{\theta} > 0$ and $k_{\alpha}$ is finite, there will be a sufficiently high $n$ for which $n\,k_{\theta} >> k_{\alpha}$ and thus $s_{11} \propto 1$ behaviour is recovered ($\frac{c}{n}$ is vanishingly small in this limit). Only in the case that $k_{\theta} = 0$, would $n = \infty$ represent a system with no resistance to changing $\theta$ and therefore have infinite in-plane compliance. Providing $k_{\theta} > 0$ and $k_{\alpha}$ is finite, all compliances behave as \nicefrac{$k_{\alpha}$}{$k_{\theta}$} $<< 1$ when $n >> 1$ and therefore in the limit $n \rightarrow \infty$, the in-plane compliance $s_{11}$ tends to a constant value whereas both axial compliance $s_{33}$ and elastic coupling between the 1 and 3 axes $s_{13}$ tend to 0 for all \nicefrac{$k_{\alpha}$}{$k_{\theta}$}.

\section{Conclusions}

For a tetragonal A$_2$BO$_4$ Ruddlesden--Popper structure with a frozen octahedral rotation about the layering axis, we assumed that the four shortest metal--anion bonds are so stiff compared to other interatomic interactions that these bonds remain perfectly rigid. This assumption leaves only a single internal degree of structural freedom, the rotation angle, $\theta$, which is directly coupled to the in-plane and axial lattice parameters. For this model, the previously proposed ``corkscrew" mechanism, we formulated equations relating structural parameters to $\theta$ and the bond lengths of the four fixed bonds. Analysing these equations allowed us to identify the structural limitations of the model and considering the parameter values at which the assumptions of the model might break down allowed us to identify a ``physically relevant" region of the parameter space. Assuming that the only resistance to deformation comes from resistance to change in bond angles, we were further able to derive equations for components of the elastic compliance matrix and investigate how these compliances behave with changes in $\theta$, the ratios of fixed bond lengths and the relative strengths of the harmonic potentials on the bond angles. Finally, we extended this model for the $n = 1$ member of the A$_{n+1}$B$_n$O$_{3n+1}$ series to a general $n$ and discussed how different compliance components change with $n$. Although this idealised model is unlikely to quantitatively represent any real chemical system, analysing the corkscrew mechanism in this limit might help understand real systems where the mechanism operates in addition to other physical effects.

\section{Acknowledgements}

CA is supported by a studentship in the Centre for Doctoral Training on Theory and Simulation of Materials at Imperial College London funded by the EPSRC (EP/L015579/1). This work was supported by the Thomas Young Centre under grant TYC-101. CA acknowledges support from the Armourers and Brasiers Gauntlet Trust for a research studentship travel grant. MS acknowledges the Royal Society for a fellowship.


\end{document}